# A Novel Methodology to Overcome Routing Misbehavior in MANET using Retaliation Model

Md. Amir Khusru Akhtar[1] and G. Sahoo[2]

[1]Department of CSE, Cambridge Institute of Technology, Ranchi, Jharkhand, India
`akru2008@gmail.com`
[2]Department of IT, Birla Institute of Technology, Mesra, Ranchi, India
`gsahoo@bitmesra.ac.in`

## ABSTRACT

*MANET is a cooperative network in which nodes are responsible for forwarding as well as routing. Noncooperation is still a big challenge that certainly degrades the performance and reliability of a MANET. This paper presents a novel methodology to overcome routing misbehavior in MANET using Retaliation Model. In this model node misbehavior is watched and an equivalent misbehavior is given in return. This model employs several parameters such as number of packets forwarded, number of packets received for forwarding, packet forwarding ratio etc. to calculate Grade and Bonus Points. The Grade is used to isolate selfish nodes from the routing paths and the Bonus Points defines the number of packets dropped by an honest node in retaliation over its misconducts. The implementation is done in "GloMoSim" on top of the DSR protocol. We obtained up to 40% packet delivery ratio with a cost of a minimum of 7.5% overhead compared to DSR. To minimize total control traffic overhead we have included the FG Model with our model and it reduces the overhead up to 75%. This model enforces cooperation due to its stricter punishment strategy and justifies its name.*

## KEYWORDS

*Retaliation, Grade, Bonus Points, Neighbor, Cooperation*

## ABBREVIATIONS

*Bonus Points (BP), Local Bonus Point (LBP), Grade (G), Number of Packet Forwarded (NPF), Number of Packet Received for Forwarding (NPRF), Packet Forwarding Ratio (PFR), Mobile Adhoc Network (MANET), Misbehavior Gain (MG), Promiscuous Mode Off (POFF), Promiscuous Mode On (PON), Plain DSR (PDSR), Modified DSR (MDSR), Friendly group with Modified DSR (FGMDSR)*

## 1. INTRODUCTION

MANETs are the cooperative networks that work without any fixed infrastructure or access point. To impel the correct functioning of the Adhoc network it is actually more difficult than the wired network because of the imprudent design. Attacks and misbehavior are certainly the wall that obstructs the development and implementation of the MANET. In this paper we are using misbehavior term for packet dropping attacks or selfishness. A selfish node does not cooperate in network participation to save its battery lifetime and bandwidth. Existing mechanisms protect at somehow but still faces other challenges such as battery lifetime and bandwidth. Proposing and establishing a secured, reliable and applicable design that suits all applications is still a big challenge for all. To model the network we have to guarantee that the model not only secures the network but also provides minimum battery usage, reliability and throughput for which the MANET was actually designed.

We are proposing a model to overcome misbehavior from a network by enforcing cooperation using the stricter punishment design called "Retaliation Model". In this model node's behavior is watched by its neighbors in promiscuous listening mode, to update the NPF/NPRF value for a specified time. After the expiry of the defined time every node calculates the PFR value and broadcasts in its neighborhood. Finally all neighbors broadcasted PFR values is received and processed by the nodes to define the 'G' and 'BP' values. The Grade is used to isolate selfish nodes from the routing paths and the Bonus Points defines the number of packets dropped by an honest node in return of selfish neighbor misbehavior.

In this model each node has to maximize the PFR up to 100% as it is used to define the grade. Therefore, for every packet loss a node will be punished and the punishment cost is in terms of its packet drop by the entire neighbors. The selfish nodes are punished by honest nodes by dropping packets intended for, or originated from, such a node. The new route is defined on the basis of Grade by bypassing such misbehaving nodes. A node is punished till the BP is greater than zero and after that the selfish node is automatically added to the network because the positive BP value denotes its selfishness. This model does not use any kind of elimination and addition algorithm because a node is punished on the basis of BP which minimizes battery usage.

In a MANET nodes become selfish because of its limited resources (such as battery power and bandwidth), that is why the packet dropping behavior or selfishness would take place. To prevent from the selfishness we have defined BP that denotes the amount of packets to be dropped by an honest node against a selfish node in retaliation over its misconducts. Thus, a node who wants to save its resources must know that by dropping packets of others, it has to spend more energy to rebroadcast the same packet again and again. Because, its packets are dropped by all its neighbors till the BP value in each node is greater than zero. The punishment cost is substantially more than to act like an honest node because more energy will be needed to rebroadcast the same packet. So, a selfish node knows that selfishness will be harmful, and will be forced to be cooperative.

Use of Retaliation Model over DSR [16] protocol to overcome the misbehavior can also be used to enhance the DSR protocol by overhearing any communications within its neighborhood. A route reply (RREP) packet can be snooped and a new source route can be added to its route cache. This would minimize the routing overhead incurred due to initiating a route request in further routing. We have deliberately not incorporated this concept in this paper.

This model gives the chances of saving energy to honest nodes by dropping packets of selfish/misbehaved nodes as well as enforces stricter punishment strategy. This model ensures cooperation and reliability in MANET because rather than eliminating it behaves in the same way as the node behaved. Therefore, it justifies its name.

The rest of the chapters are organized as follows. Section 2 presents background and existing work on misbehavior detection and prevention. Section 3 presents the Retaliation Model. Section 4 gives the simulation specific assumptions, simulation results and discussion of the results. Section 5 concludes the paper.

## **2. BACKGROUND AND RELATED WORK**

Lots of attacks based on modification of routing data can be handled using secure routing protocols [10-14] but when nodes show its non cooperation or selfishness all these protocols fails.

To enforce cooperation several methods [19-41] have been proposed but they consume more energy and bandwidth which are the actual cause of non cooperation.
Our model is a reputation based model that is why we are concentrating on the detection of misbehavior using 'Reputation Based Mechanisms'.

Detection of routing misbehavior was first proposed by Marti et al. [1] using Watchdog and Pathrater. This mechanism was proposed to be used over the DSR [16] routing protocol to mitigate routing misbehavior (including selfish nodes) in ad hoc networks. The watchdog is based on neighbor monitoring for identifying malicious and selfish nodes while Pathrater evaluates the overall reputation of nodes on a path. Pathrater defines a route by excluding selfish nodes or misbehaving nodes lying on the paths. In this mechanism selfish nodes are rewarded because there is no punishment for the same. It has another serious drawback that extra battery power consumption because every node has to constantly listen to the medium.

Buchegger et al. [2, 3] introduced the CONFIDANT protocol to observe the behavior of nodes, calculate the reputation of corresponding nodes, and punish the identified selfish nodes. CONFIDANT protocol has four parts: a monitor, a reputation system, a trust manager and a path manager. The Monitor is responsible for recording the behavior information of neighboring nodes. The reputation system is responsible for calculating the reputation of nodes on the basis of direct observation and friends' (indirect) observation. The trust manager is defined to collect warning messages from friends, and the path manager is used to manage routing by excluding selfish nodes. In this protocol, each node monitors its neighborhood behavior and observed misbehavior is reported to the reputation system. If the misbehavior is intolerable then it is reported to the path manager, and then the path manager excludes the nodes from the routing path and calculates new paths. In this regard a warning message will be sent from the trust manager to the friends regarding the misbehaved nodes and after receiving the warning message, the trust manager of a receiver computes the trustworthiness of the message and passes it on to its reputation system if necessary. CONFIDANT has weaknesses in terms of an inconsistent evaluation problem, because in this system each node evaluates different evaluations for the same node and therefore, it is difficult to identify a selfish node. It has another weakness of a location privilege problem because it punishes on the basis of packet dropping not on the basis of how they contributed to the network before. This will drain more battery power for a node situated in the center of the network than the nodes lying on the periphery of the network.

Michiardi et al. [4] proposed reputation measure to know a node's contribution to a network. Reputation is classified into three types: subjective, indirect and functional. Subjective reputation is computed on the basis of node's direct observation, Indirect reputation is computed based on the information provided by other nodes and Functional reputation the subjective and indirect reputation with respect to different functions. It concentrates only routing function and packet forwarding function. After that is takes these reputations to aggregate a collaborative reputation.
Michiardi et al. suggested the CORE [5] protocol to evaluate nodes on the basis of three reputations i.e., collaborative reputation. In this work it maintains a set of Reputation tables at each node with a watchdog mechanism. The watchdog mechanism is used to observe whether a required function is correctly performed by the requested node by comparing the observed execution of the function with the expected result. The Reputation table is used to maintain the reputation value of the nodes in the network. Reputation is created/updated along time, based on direct listening by the node itself, or on the basis of information provided by others. This will help a node to judge the selfishness of a service requester and thus decide either to serve or to refuse the request.

Miranda et al. [6] suggested that a node periodically broadcast information's about the status of its neighboring nodes and nodes are allowed to globally declare their refusal to forward messages to certain nodes. This mechanism gives higher communication overhead.

Paul and Westhoff [7] proposed security extensions to the existing DSR protocol to detect attacks in the process of routing. The mechanism depends on neighbor's observations and the routing message's in order to detect the attacker.

Friends & Foes [8] is a mechanism in which Friends receives the services of a network and Foes refuses to serve by the nodes. This mechanism works fine but with memory and message overhead. In RIW [9] emphasis is given on current feedback items relatively than old ones. It keeps a node behaving selfishly for a long run to build a good reputation. This approach is good but not a practical solution.

To enforce cooperation lots of methods [19-41] have been proposed but they consume more energy and bandwidth which are the actual cause of non cooperation.

The Friendly Group [42] model proposed by Akhtar & Sahoo is an approach for securing an adhoc network by involving two Network Interface Cards (NICs) in each node to partition a MANET into several friendly groups/subnets. This model enhances cooperation by minimizing battery usage but it is not a suitable solution for all applications.

## 3. PROPOSED WORK

In this section we have presented the "Retaliation Model" to enforce co-operation and eliminate misbehavior that gives a secured and reliable platform to execute MANET. It can be implemented on the top of existing DSR protocol [16].

### 3.1. Overview

MANETs may be considered as a society in which nodes agree to co-operate with each other to fulfill the common goal, but non-cooperation is genuine to save itself in terms of their battery power and bandwidth. As we know, cooperation is the basic requirement of MANET that is why we are defining this model that strictly enforces cooperation and eliminates misbehavior. We have defined a grading system to quantify the selfishness of a node. We have used packet forwarding ratio (PFR) as criterion where PFR is the ratio of the number of packets forwarded (NPF) to the total number of packets received for forwarding (NPRF) and it shows a node's contribution to the network.

Every DSR node implements an instance of the 'Retaliation Model' and runs in two modes as given below. We have used GlomoSim, a scalable network simulator [18], to simulate our work. At the start of the simulation the MANET must run in protected mode to obtain the Grade and Bonus points to initiate cooperative start.

Normal Mode or POFF mode: in this mode node does not overhear, only regulates its normal activities. It routes packets according to the previous G and BP values on top of the DSR routing protocol. A node has to use Grade and Bonus Points in the route request, route reply and route error activities. G value is used to forward a packet by isolating the low grade nodes from the routing paths and BP is used to punish selfish neighbor by dropping its packets. A node has to drop the BP amount of packets that results in saving of energy as well as punishing selfish nodes. Modification of NFR and NPRF values are performed in the Protected Mode.

Protected Mode: In this mode node overhear as well as handle/routes packets according to G and BP values on top of the DSR routing protocol. Nodes are supposed to be in an observation mode in which all nodes must behave well to obtain a higher grade. Here we are capturing the parameters for the node such as number of packets received for forwarding, number of packets forwarded at each node in the neighborhood. These values can be processed to obtain the packet forwarding ratio that will be used to calculate the Grade and Bonus Points.

The proposed model contains three modules i.e. Neighborhood Behavior Detection Module, Processing, Grading and Bonus Points Allocation Module and Punishment Definition Module. The system architecture is given in Figure 1.

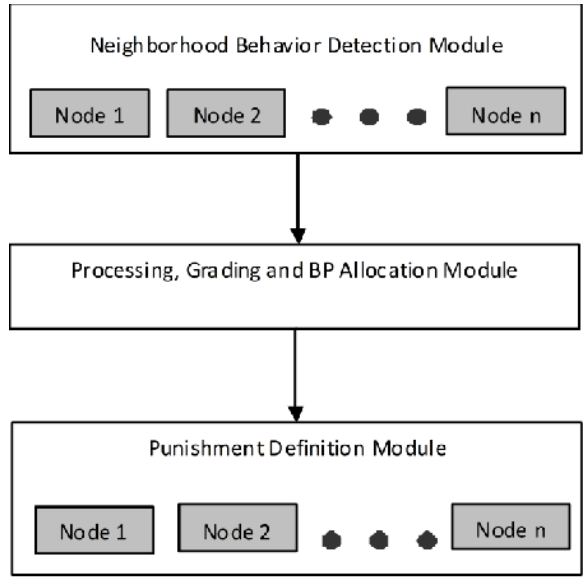

Figure 1. Retaliation Model

### 3.2. Elementary terminologies

Definition 1. Grade (G): it quantifies the selfishness of a node in its neighborhood.
Definition 2. Bonus Points (BP): it denotes the amount of packets to be dropped by an honest node against a selfish node in spite of it misbehavior/packet drops.

### 3.3. Description of modules

These modules are defined as follows

### 3.3.1. Neighborhood behavior detection Module

This module monitors the neighborhood behavior by promiscuous listing the neighbor traffic. The Retaliation process runs on each node for getting information about the neighborhood. It stores the behavior information into a table; NI table (Neighborhood Information Table). This table contains a unique entry for each node of the neighborhood. Nodes have to update the NPRF and NPF values on the basis of number of packets received for forwarding and number of packets forwarded. The schema of the NI table is given below:

191

NI (IP, NPRF, NPF, G, BP)
where  IP:    Internet Protocol Address
          NPRF:  No of Packet Received for Forwarding
          NPF:    No of Packets Forwarded
          G:       Grade
          BP:     Bonus Points

IP field defines the identity of the node in the MANET. Each node monitors its neighborhood behavior by promiscuous listening to increment the NPRF/NPF value of its neighbor. Before incrementing these values a node has to check the BP and if it is greater than zero it shows misbehavior of the corresponding node. The listening node decrements the BP after a packet drop/ allow to drop from the corresponding neighbor and when it is equal to zero then only the increment of the NPRF/ NPF value is performed.

NPRF, NPF, and BP is initialized with zeros and G is initialized with one. The zero value of the given fields indicates the fresh start of the simulation/MANET. The grade is initialized with one because this model assumes that a new unknown node is honest. The NPRF and NPF values will be subsequently updated by overhearing the neighbors on the basis of number of packets received for forwarding and number of packets forwarded.

The Neighborhood Behavior Detection Module working diagram is given in Figure 2. In which we have shown 15 nodes arranged in a grid pattern. The nodes, updates its NI table in promiscuous mode, an instance of NI table (for node B) is also shown in Figure 2.

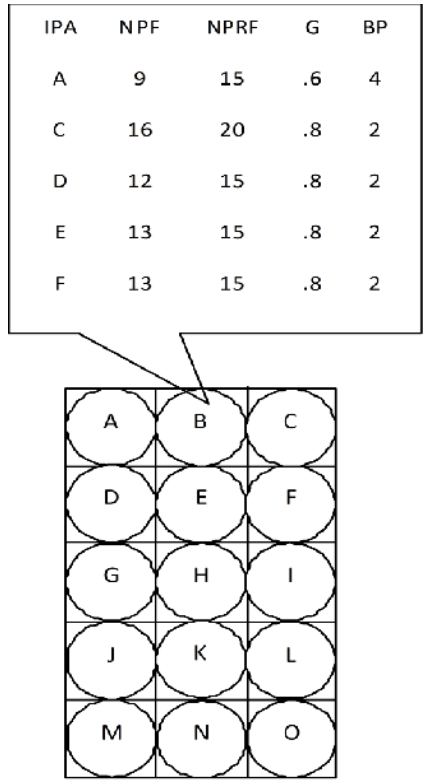

Figure 2. Neighborhood Behavior Detection

International Journal of Wireless & Mobile Networks (IJWMN) Vol. 5, No. 4, August 2013### 3.3.2 Processing, Grading and BP Allocation Module

After updating the NI table for the threshold time in the promiscuous listening mode, every node has to process the NI table. At first a node calculates the PFR values for its neighbors and broadcast it along with the IP Address to its neighbors. The formula to calculate the PFR is
PFR = No of Packet Forwarded / No of Packet Received for Forwarding
Similarly every node broadcasts its neighbors IP and PFR that will be accepted and filtered (only neighbor information) by the node. This information is kept in a Temp table which is defined as follows.

　　IT (IP, PFR, G, BP)

| IP | PFR | G | BP |
|----|-----|---|----|
| A | 0.6, 0.8, 0.7 | 0.7 | 3 |
| C | 0.8, 0.7, 0.9 | 0.8 | 2 |
| D | 0.8, 0.7, 0.7, 0.8, 0.6 | 0.7 | 3 |
| E | 0.8, 0.5, 0.7, 0.8, 0.9, 0.9, 0.8, 0.4 | 0.8 | 2 |
| F | 0.8, 0.8, 0.7, 0.7, 0.8 | 0.8 | 2 |

Table 1. Temp table for Node B

Where PFR and BP are multivalued attributes, it stores PFR and BP values received from its neighbors. First of all a node writes its own PFR value in the PFR cell and then it appends this field by the received PFR values from its neighbors. An instance of Temp table is shown in Table 1.

After storing the relevant information in the Temp table nodes have to calculate the Grade and Local Bonus point (LBP). The steps are as follows.

Step1: [Calculation of Grade for the i[th] node]

　　The grade is obtained by calculating Mean for the PFR values as

$$G_i = \sum_{k=1}^{n} PFR_k / n$$

　　Where n defines the number of neighbours

Step 2: [Assigning Local Bonus Points for i[th] node]

$$LBP_i = 2^{(MG_i * 10)}$$

The LBP is calculated and stored in BP cell locally (in which a node assigns bonus point to its neighbors on the basis of Grade). We are subtracting G from 1 (because 0 ≤ G ≤ 1) to find the MG (Misbehavior Gain).

where $MG_i = 1-G_i$

MG is multiplied by ten to make it an integer value that easily defines how many packets to drop for node i and it is increasing exponentially to enforce strict cooperation.

### 3.3.3 Punishment Definition Module

The punishment will be defined on the basis of LBP. After calculating G and LBP for its neighborhood a node has to broadcast the LBP along with the IP address to its neighbors and similarly all broadcasted LBP of its neighbors would be appended in the BP cell of the Temp table. Then the mean value of the LBP is calculated that will decide how many packets will be dropped by an honest node against a selfish node in spite of its misbehavior/packet drops. We have calculated the mean value for the LBP cell to make BP value consistent in a neighborhood. Punishment module is defined as follows

Step 1: [Calculation of BP for the i$^{th}$ node]

$$BP_i = \sum_{k=1}^{n} LBP_k \ / \ count \ ([LBP])$$

Where count ([LBP]) defines the number of values in the respective LBP cell.

Step 2: [Updating of NI table]
After calculating BP the G and BP column of NI table will be initialized/updated by Temp table. The G value will help to forward a packet by isolating the low grade nodes from the routing paths and BP value is used to punish selfish neighbors by dropping BP amount of packets. This will save energy of the honest nodes as well as enforce stricter punishment strategy.

## 3.4 Adding a new node in the network

In this model inclusion of a new node is very simple, the NPRF, NPF, and BP values are initialized to zero and G is initialized with one. The zero values in the given fields indicate a fresh start of the node in network activities and the value one indicates that our model assumes an unknown node is honest. The NPF and NPRF values will be subsequently updated in protected mode by overhearing the neighbors on the basis of the number of packets forwarded or received for forwarding.

## 3.5 Combining Retaliation Model with existing Routing Protocols

This model can be simply implemented on top of the existing routing protocols such as DSR [16] and AODV [17]. In DSR protocol, when a node has a packet to send and the destination is unknown, it will broadcast a route request (RREQ) packet to its neighbors. Neighbor nodes have to append their own address in the RREQ packet and rebroadcast the RREQ packet to their neighbors. Finally the RREQ packet is received at the destination and the destination node will send a route reply (RREP) packet to the source by reverse path.

In order to combine DSR routing protocol with our scheme, packets are handled on the basis of G and BP values. If a RREQ packet arrives from the node whose BP value is greater than zero then it shows misbehavior of the corresponding node. The packet is dropped by the receiving node and therein it decrements the BP by one every time, until it is zero. In the case of overhearing, if a node is watching the traffic of other node then it has to decrement the BP by one after a packet drop from the corresponding neighbor and when it equals to zero then only it increments the NPRF/ NPF value. In case of RREQ the source includes isolate low grade nodes from the routing paths.

In AODV, whenever a source node needs a route to a destination node, it floods the network with route request RREQ packets. An intermediate node has to reply if it knows a fresh route to the destination, otherwise it propagates the request and nodes update their routing table with a reverse route to the source. When the RREQ reaches the destination, destination replies by sending a RREP towards the source with the reverse route. In the process of route maintenance, upon detecting a link break, a node sends RERR with the active route(s) towards the source(s).

We can combine this with our scheme, by the involvement of G and BP values. A node needs to check the Grade and Bonus Points whenever it gets any RREQ packet. It can drop the BP amount of RREQ packets of the misbehaved nodes. During route reply and maintenance the same punishment strategy can be applied. The source can exclude low grade nodes to initiate a new route request. Similarly, if selfish node has to send a RREQ, then it has to spend more energy because its packet has been dropped by its neighbor till BP reaches zero.

## 3.6 Minimization of total control traffic overhead by dividing a MANET into several friendly groups

We can minimize the total control traffic overhead by dividing a MANET into several friendly groups suitable for some applications. In FG Model [42] a network is divided using several friendly groups using two NICs installed in each node. The proposed diagram in Figure 3 shows the proposed Friendly Group structure. In which the four different FGs are indicated by cross, triangle, circle and square.

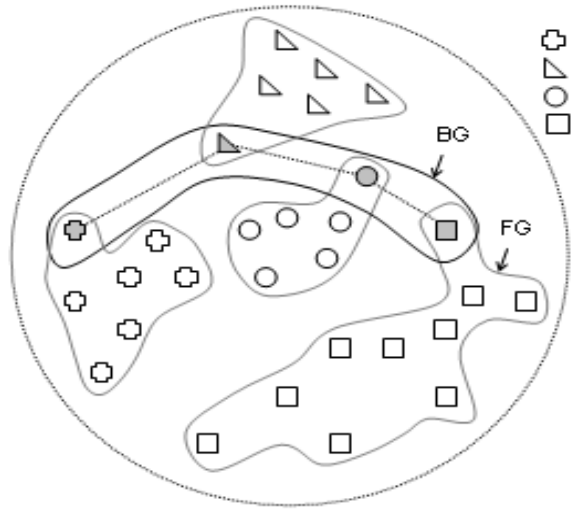

Figure 3. Friendly Group Architecture of four FG with one BG

where, FG denotes Friendly Group and BG denotes Border Group as defined in [42].
The advantage of the inclusion of FG model with our model is that it minimizes the battery usage and thus enhances the network cooperation. The nodes have fewer chances for misbehaving because they have enough energy to survive. The FG Model gives a reduction in control packets

International Journal of Wireless & Mobile Networks (IJWMN) Vol. 5, No. 4, August 2013as it divides a MANET of size N into k friendly groups with an approx N/k number of nodes per group. The total control overhead defined for reactive routing protocols [43] is $N^2$ with a flat structure and in FG (Hierarchical) structure it is $N^2/k$. This will enhance the network throughput of our proposed model up to a ratio of [k : 1 | k > 1]. The simulation result shows the benefits of reduction in overhead after inclusion of FG model with our model as defined in section 4.3.3.

## 4. SIMULATION RESULTS

In this section we discuss the details of the simulation and results.

### 4.1. Details of Simulation Environment

In this work we have used GloMoSim, a scalable network simulator [18] for simulating our Retaliation Model to overcome misbehavior in MANET. We have taken following parameters in our simulation given in Table 2.

Table 2. Simulation Parameters

| Parameters | Values |
|---|---|
| SIMULATION-TIME | 15M |
| TERRAIN-DIMENSIONS | (1250, 1250) |
| NUMBER-OF-NODES | 121 |
| NODE-PLACEMENT | GRID |
| MOBILITY | RANDOM-WAYPOINT |
| MOBILITY-WP-PAUSE | 30S |
| MOBILITY-WP-MIN-SPEED | 0 |
| MOBILITY-WP-MAX-SPEED | 10 |
| MOBILITY-POSITION-GRANULARITY | 0.5 |
| PROMISCUOUS-MODE | YES |
| ROUTING-PROTOCOL | DSR |

### 4.2. Energy Consumption

In a Mobile Adhoc Network, nodes have limited resources (such as battery life and bandwidth), that is why energy consumption becomes a major concern. The transmitted power is the strength of the emissions measured in Watts (or milliWatts). A high transmit power will drain the batteries faster, and sensitivity is the measure of the weakest signal that may be reliably heard on the channel by the receiver; the lower value of the signal depends on the receiver hardware performance. Normally values are around -80 dBm, in this model we are using the lowest -90 dBm which result better and we are assuming that normally hardware is better in MANET. In this model to minimize the battery consumption we are reducing the transmitter power because a node can broadcast or overhear its neighborhood only. We have taken the following parameters in GlomoSim for Setting up the Transmission Range is given in Table 3.

Table 3. Radio Parameters

| Parameters | Values |
|---|---|
| PROPAGATION-LIMIT (dBm) | -111 |
| PROPAGATION-PATHLOSS | Two-Ray |

196

| RADIO-FREQUENCY (Hz) | 2.40E+09 |
|---|---|
| RADIO-TX-POWER (dBm) | 1 |
| RADIO-ANTENNA-GAIN (dBm) | 0 |
| RADIO-RX-SENSITIVITY (dBm) | -91 |
| RADIO-RX-THRESHOLD (dBm) | -81 |
| RADIO RANGE (M) | 125.227 |

### 4.3. Simulation Results

To examine the system we have conducted various tests after implementing our scheme on GloMoSim simulator. Our underlying protocol for this work is DSR [16], implemented as plain DSR (PDSR) i.e., the original DSR and with our improvement, named Modified DSR (MDSR). We have analyzed the packet delivery ratio and overhead in our simulation study defined as follows.

### 4.3.1 Packet Delivery Ratio

The PDR with respect to percentage of selfish nodes is given in Figure 4. The graph shows that the PDR decreases as the percentage of selfish nodes increase. Our scheme 'MDSR' shows a higher packet delivery ratio (up to 40%) than the corresponding PDSR. We have calculated PDR using the given formula

PDR = Total number of packets received / Total number of packets sent

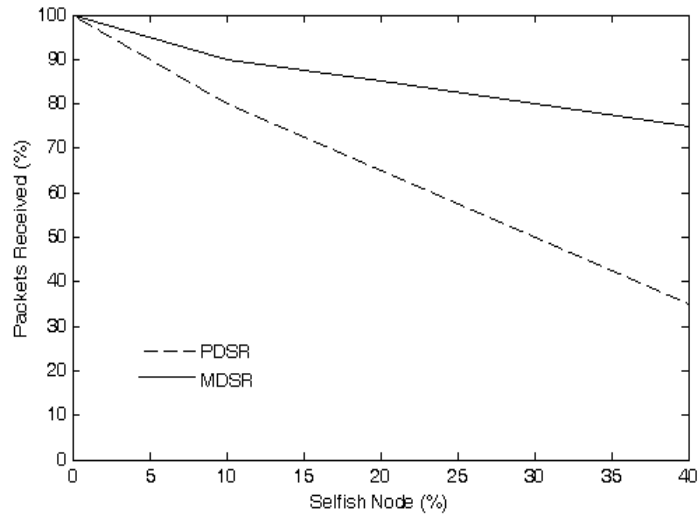

Figure 4. Packet Delivery Ratio

### 4.3.2 Overhead (Number of packets)

In Figure 5 we have shown the overhead in terms of total number of packets versus number of nodes in the network. Laurent Viennot, et al [15] proposed the control traffic overhead for reactive routing protocols, and to guarantee full connectivity in the network a node at least have to maintain a route to every other node in the network. The implementation of the Retaliation Model with DSR enhances cooperation with a cost of increase in overhead (minimum of 7.5%). The overhead decreases when the size of MANET increases.

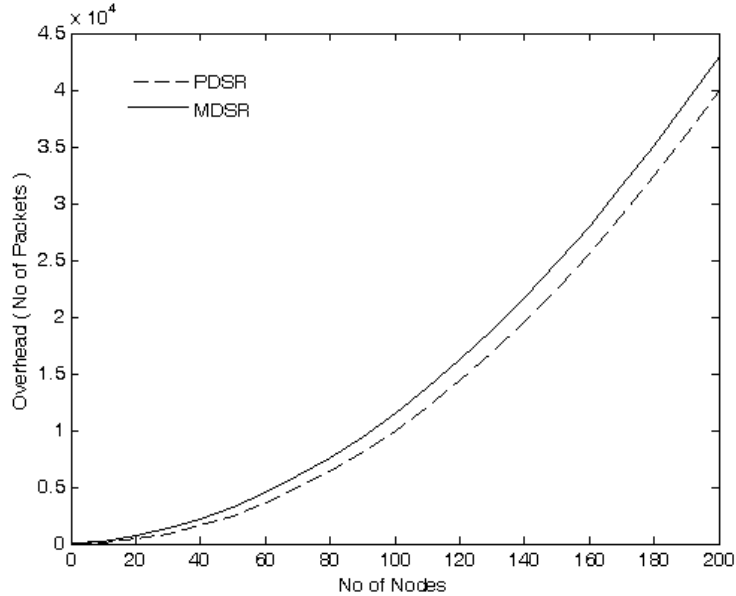

Figure 5. Overhead comparison between a PDSR and MDSR

### 4.3.3 Inclusion of Friendly Group Architecture in our Model

The inclusion of FG Model reduces the total control overhead up to a ratio of (k : 1 | k > 1). Figure 6 shows the reduction in control overhead because it divides a MANET into several friendly groups which reduces the total number of control packet transmission for full connectivity. We have denoted the Friendly Group with MDSR by FGMDSR. In this simulation we have taken four friendly groups and the obtained result shows the total reduction in overhead up to 75%. This will enhance cooperation in a MANET because nodes have more energy to survive.

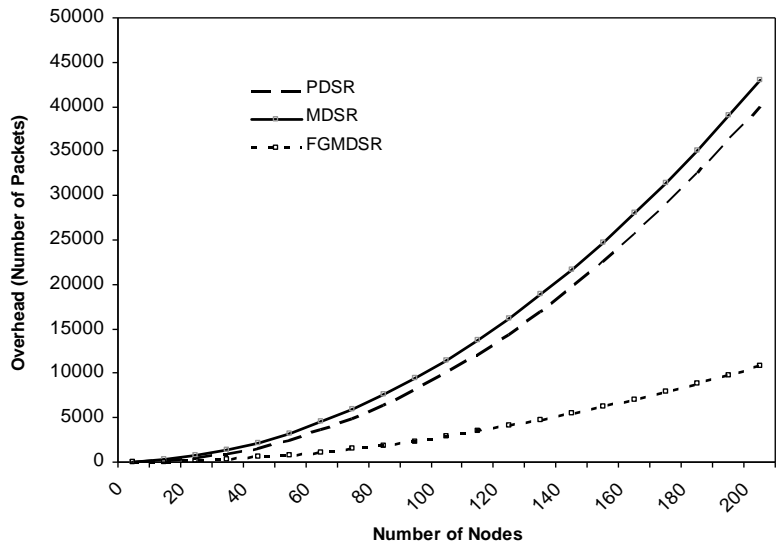

Figure 6. Overhead comparison between a PDSR, MDSR and FGMDSR

## 5. CONCLUSIONS

The presented "Retaliation Model" enhances cooperation in MANET by involving its stricter punishment strategy that is an equivalent misbehavior given in return. It uses a new method of punishment on the basis of 'Grade' and 'Bonus Points'. The Grade is used to isolate selfish nodes from the routing paths and the Bonus Points defines the number of packets dropped by an honest node in return of selfish neighbor misbehavior. Here we have enforces cooperation at neighborhood level which finally overcomes misbehavior from the entire MANET. Simulation results show up to 40% packet delivery ratio with a cost of a minimum of 7.5% overhead compared to the plain DSR. To minimize total control traffic overhead we have included the FG Model with our model and it reduces the overhead up to 75%. In addition to that we have reduced transmitter power because a node has to broadcast or overhear nodes in its neighborhood only which enhances battery lifetime also. Our model ensures cooperation and saves battery power because it does not define any complex elimination algorithm but it behaves in the same way as the node behaved. This will warn selfish nodes to cooperate in network partition because honest modes will drop packets of selfish nodes in retaliation over its misconducts. Therefore, "Retaliation Model" justifies its name.

## ACKNOWLEDGEMENTS

The authors are thankful to anonymous referees and WiMoN 2013 Team for their valued comments and suggestions to get better the quality of the paper.

**Authors**

Md. Amir Khusru Akhtar

Md. Amir Khusru Akhtar received his M. Tech. in Computer Science. & Engg. from Birla Institute of Technology, Mesra, Ranchi, India in the year 2009 and is pursuing PhD in the area of Mobile Adhoc Network from Birla Institute of Technology, Mesra, Ranchi, India. Currently, he is working as an Assistant Professor in the Department of CS & E, Cambridge I nstitute of Technology, Tatisilwai, Ranchi, Jharkhand, India. His research interest includes mobile adhoc network, parallel and distributed computing and cloud computing. 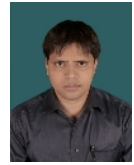

G. Sahoo
G. Sahoo received his MSc in Mathematics from Utkal University in the year 1980 and PhD in the Area of Computational Mathematics from Indian Institute of Technology, Kharagpur in the year 1987. He has been associated with Birla Institute of Technology, Mesra, Ranchi, India since 1988, and currently, he is working as a Professor and Head in the Department of Information Technolo gy. His research interest includes theoretical computer science, parallel and distributed computing, cloud computing, evolutionary computing, information security, image processing and pattern recognition. 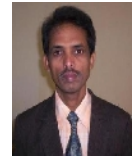